\documentclass{elsart}

\usepackage{amssymb}
\usepackage{graphics}
\begin{document}

\begin{frontmatter}

\title{The Penrose Lattice Revisited} 

\author{A. Losev}

\address{Bulgarian Academy of Sciences,
\\ Institute of General and Inorganic Chemistry,
\\ 11 G. Bonchev str., Sofia, Bulgaria }
\ead{alosev@svr.igic.bas.bg }

\begin{abstract}
A recursive scheme relying on decagons is used to generate Penrose-like sublattices or tilings. Its relevance for understanding structures with non-crystallographic symmetry is discussed.
\\pacs 61.44Br
\end{abstract}

\end{frontmatter}

\maketitle

\section{Introduction}
The discovery of quasicrystals twenty-five years ago\cite{Shecht} greatly enhanced the interest in a class of mathematical artifacts known as aperiodic tilings. The formal approach to such structures readily exhausted most of the topic\cite{Moody,Zorka} but the existence of their material counterparts is still poorly understood\cite{discuss}. The gap between a physical or a mathematical approach has been partly bridged by the concept of covering which allowed application of extremal principles to produce aperiodic tilings from overlapping  geometrical shapes\cite{Gahler}. A further development was the discovery of non-crystallographic symmetries produced by mechanical\cite{Edw} and optical waves\cite{Gorkhali} and also the existence of soft quasicrysals\cite{Lifshitz}.
A paradigmatic case for all research in this area has been the Penrose tiling (PT), its embodiment in icosahedral or decagonal quasicrystals and systems of Faraday waves with fivefold symmetry. It is invariably mentioned in almost any related publication, with various aspects discussed.

Below we investigate a construction with decagons which allows to generate sublattices  and coverings for the PT, relying on a recursive scheme. This leads to a speculation about general principles and realistic models.

\section{\bf{Model and Method}}
Lattices are discrete sets of points whose properties in some cases justify descriptions as 'aperiodic', 'nonperiodic' or 'quasiperiodic'. In two dimensions non-overlapping polygons, or 'tiles', provide an alternate description, their vertices being the the points of a lattice. A PT is a nonperiodic tiling generated by an aperiodic set of tiles discovered by Roger Penrose.  Some variants and related tilings or sublattices have been already considered\cite{Luck1}. Gummelt\cite{Gumm1} has shown that copies of a suitably decorated decagon, when allowed to overlap, can cover the plane producing a perfect PT.  The decoration is made with the so called 'kites' obtained from pairs of golden triangles. As a local approach this construction relies on the two  kinds of overlap which are allowed.
The use of decagons, rather obvious in this context, has been previously commented and reconsiderated \cite {Shenbu} and the present investigation also relies on it. The hierarchical structure of the PT in fact contains a  recursive scheme: an originary decagon produces successive generations of ten decagons arranged in a regular ring. The ring radius is increased by the golden mean $\phi$ at each generation while the decagons from a previous generation spawn just one new generation. Due to the geometrical properties and scaling many of the vertices generated in this way coincide. It is easy to check that the first ring contains just 50 different points and indeed the proliferation is much more limited as it might be thought. Also  just two kinds of overlap arise. This whole construction possesses a tenfold rotational symmetry and could be considered to be the 'backbone' of the tiling. The productive step appears to be a simple modification of the decagons: at each vertex an outward pointing 'spike' is attached. 

%[Fig.1 here]
%{\small A piece of cogwheel tiling. The center of rotational symmetry %is just below the frame, to the right of its middle.}

\begin{figure} 
\resizebox{0.85\textwidth}{!}{%
\includegraphics{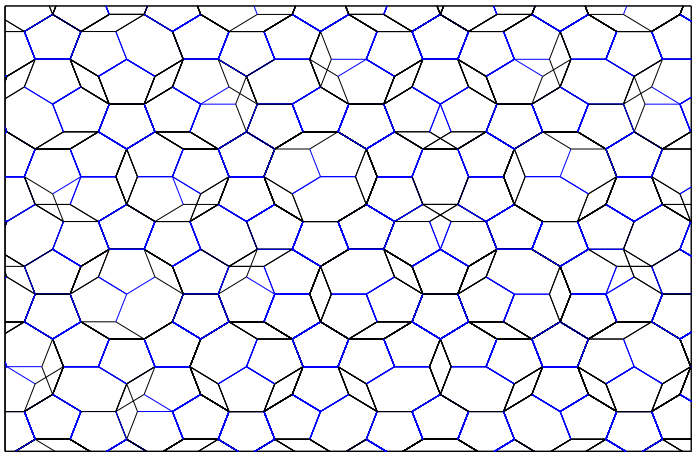}
}
\caption{A piece of cogwheel tiling. The center of rotational symmetry is just below the frame, to the right of its middle.}
\label{Fig.1}      
\end{figure}

 A patch from the resulting tiling is shown in Fig.1. To distinguish it from an earlier related proposal\cite{al} we would call this a cogwheel tiling. Its basic element consists of twenty points arranged in a double decagonal ring whose width is equal to the inner decagon's side and it is drawn as a spiky decagon or a cogwheel. Even if it is not a proper tile, a simple tessellation of the plane is obtained.

\section{\bf{ Results and Discussion}}

The resemblance, or the difference, of this tiling with  the original PT, known as P1, is more or less obvious. In fact the set of vertices defining the edges drawn here, is a subset of the Penrose lattice. In such cases it is dubious how much the linking of all equidistant pairs of points clarifies or obscures the result. For instance the so called 'boats' of P1 here appear crossed by two spurious lines. However the outlined shapes exhibit the major difference: P1 is built with four shapes - a pentagon, a thin rhombus, a boat (3/5 of a star) and a fivefold star. The first three of these shapes are present here but the fourth is an elongated hexagon. Adding a new point inside the hexagon allows to implement the reconfiguration:

{\it elongated  hexagon + point + boat = pentagon + star}.

So the hexagons outline missing points from the P1 lattice. With a finite number of steps, towards the periphery, incompletely  tessallated decagons appear in increasing numbers. To decompose a decagon three points are need (the center being the fourth), and some of the flaws observed in the tiling are due to decagons with one missing point inside. There are two possibilities for its position and if both are exploited a crossed boat would appear.

%[Fig.2 here]
%{\small A piece of cogwheel tiling with fivefold rotational symmetry.}

\begin{figure}
\resizebox{1.25\textwidth}{!}{%
\includegraphics{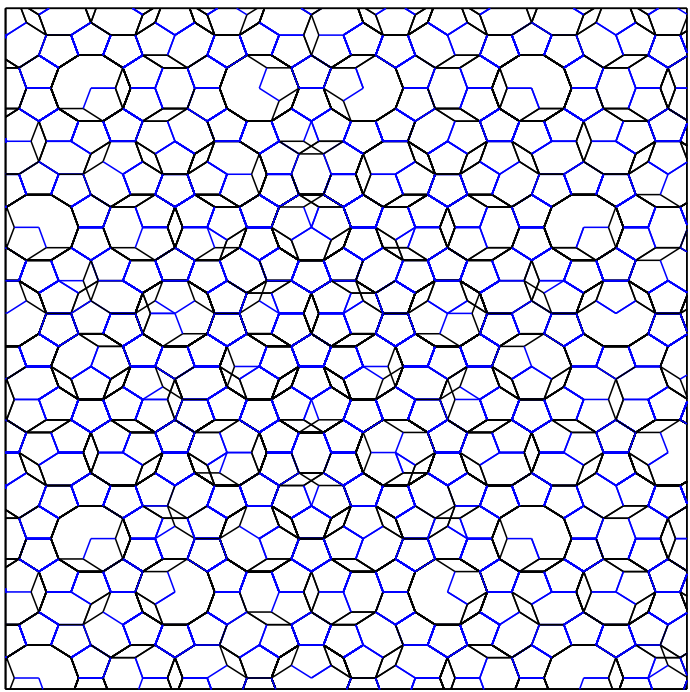}
}
\caption{A piece of cogwheel tiling with fivefold rotational symmetry.}
\label{Fig.2}      
\end{figure}

 At the cost of a slight complication it is possible to apply the same recursive construction at a fivefold origin: the first ring of five overlapping decagons appears around a pentagon. The following generations appear at  distances increasing  as $\phi^2$ and rotated by $\pi/10$. These originary decagons breed as described.  A piece of cogwheel  tiling around the global fivefold origin is shown in Fig.2. The tenfold case can be considered to occur indefinitely far from this global center of fivefold symmetry. 
Some flaws in the tiling remain but they are the price for the great overall simplification: one shape, the spiky decagon or cogwheel, is used and recursion is the only rule.

A previous work introduced the spiky decagon tiling (SDT), grown by local rules and restricted by symmetry. It is a more perfect construction which uses just three pieces, a pentagon, a rhombus and the elongated hexagon, but it is obtained through local forced and unforced steps for adding new decagons. The same tessellation is produced but only one kind of overlap was allowed: two decagons share a rhombus only when their hexagons share a vertex. The recursion now imposes another kind of overlap over hexagons and which leads to the crossed boats. Indeed the two kind of overlap, a small and a large one, are known from Gummelt's work. The interesting result of this investigation appears to be the recursive backbone of the these tilings. The SDT was shown to cover a PT \cite{al} and the same can be done here. On a P1 tiling the decagons can be drawn and where stars are crossed, the smaller fragment (a 'crown' or 2/5 of a star) is reunited with the wedged pentagon to produce an elongated hexagon. The vertices of the cogwheel tiling, being a subset of P1, are  obtained by discarding a suitably chosen point from each star. However, in the absence of a rule, adding the 'missing' points in a cogwheel tiling would produce some imperfect or random PT.

Weighting known facts about structures with non-crystallographic symmetry might show that it is the existence of quasicrystals which is an oddity and not the patterns in Faraday waves. Oscillations with incommensurate periods seem to offer  motivated descriptions or models and suggest to look for other connexions. Indeed the Huygens principle implies recursion and central symmetry, so the twin facts that many aperiodic tilings possess a hierarchical structure and that they can be constructed using regular polygons\cite{Gahler} could hint eventually at some kind of a discrete equivalent\cite{Enders}.  Indeed for $3-, 4-$ and $6-$ fold cases a  trivial  recursion mimics the principle, allowing to cover the plane (and even the propagation in presence of obstacles). In the five- and ten -fold  cases the  recursion occurs within a geometrical progression but the property of the golden mean

$\phi^k=f_{k-1}\phi+f_{k-2}$

where $f_k$ is the $k-$th Fibonacci number, allows to see each new value as arising through a sum of previous ones.
A preceding attempt to exploit incommensurate waves failed to produce perfect nonperiodic structures but suggested to reverse the perspective: in some cases perfection might  be just an artifact which makes physical structures appear as 'flawed'. 

\section{\bf{ Conclusion}}
A recursive approach revealed a backbone built from decagons present in Penrose-like structures. Two kinds of overlap occur between pairs of basic 'cogwheel' elements defined with just one unit length. The obtained might be somewhat imperfect but its missing points mark ambiguous locations which correspond to possible  phason flips. The construction hints perhaps at a more close relationship between waves and quasicrystals.

\section{\bf{List of Figures and Captions}}
Fig1. A piece of cogwheel tiling. The center of rotational tenfold symmetry is just below the frame, to the right of its middle.
\\
Fig2. A piece of cogwheel tiling with fivefold rotational symmetry.

\end{document}